\documentstyle[aps,epsf,preprint]{revtex}
\input{epsf}
\begin{document}
\draft
\title{
Validity of the Color Dipole Approximation for Diffractive Production of 
Heavy Quarkonium}
\author{K.~Suzuki$^1$\footnotemark[1], A.~Hayashigaki$^1$\footnotemark[2], 
K.~Itakura$^2$\footnotemark[3], 
J.~Alam$^1$\footnotemark[4], T.~Hatsuda$^1$\footnotemark[5]}
\address{$^1$Department of Physics, University of Tokyo, Tokyo 113-0033, 
Japan}
\address{$^2$Radiation laboratory, RIKEN, Wako, Saitama 351-0198, Japan}
\footnotetext[1]{e-mail: ksuzuki@nt.phys.s.u-tokyo.ac.jp}
\footnotetext[2]{e-mail: arata@nt.phys.s.u-tokyo.ac.jp}
\footnotetext[3]{e-mail: itakura@rcnp.osaka-u.ac.jp}
\footnotetext[4]{e-mail: alam@nt.phys.s.u-tokyo.ac.jp}
\footnotetext[5]{e-mail: hatsuda@phys.s.u-tokyo.ac.jp}

\maketitle
\begin{abstract}
We study the diffractive photo- and leptoproductions of 
$J / \psi$ and $\psi '$ on the proton, and examine the validity of the 
small-size color 
dipole approximation to the production of radially excited heavy quarkonium.  
The dipole model predicts a small
ratio of $\psi '$ to $J / \psi$ photoproduction 
cross sections, which does not agree with experimental data.   
We show that this discrepancy originates from a large transverse size 
of $\psi '$ which makes the convergence of the transverse size expansion 
questionable, 
and the calculation without the dipole approximation 
turns out to be consistent with the data.  
Productions of  $\Upsilon '(2S,3S)$ are also discussed, and 
the dipole approximation is found to be 
reasonable for the $\Upsilon$-family.  
\end{abstract}

\pacs{13.85.Ni,14.40.Gx,12.38.Bx,12.39.Jh}



\section{Introduction}

Recently diffractive phenomena in QCD have attracted 
considerable theoretical and experimental interests.  
In particular, diffractive photo- and leptoproductions of vector 
mesons off the proton, $\gamma^{(*)} + p \to V(\rho^0, \omega, 
\phi,J/\psi, \dots) + p$, provide crucial constraints on 
properties of the QCD Pomeron and the vector meson wave 
functions\cite{Ryskin,Brodsky,FKS,Ryskin2,Martin,Nemchik,DL,Review}.  
In this report, we focus on the heavy quarkonium production, where the 
heavy quark mass plays a role of the hard scale.    
Due to the Regge kinematics, the vector meson production 
can be understood  as the scattering of the color dipole with the 
target nucleon.  
In this case we expect the amplitude to be dominated by the four diagrams 
illustrated in Fig.1.  
The upper part of each diagram, where the heavy quarks interact with 
two-gluons, is assumed to be calculated within perturbative QCD
even for the photoproduction.  
On the other hand, 
the lower blob represents both non-perturbative structure (impact 
factor) of the proton and complicated gluon dynamics at small-$x$ 
which is identified with the QCD Pomeron.

The color dipole  approach is adopted to evaluate the $J / \psi$ production 
in refs.~\cite{Brodsky,FKS,Nemchik}, 
and gives reasonable results in agreement with the data.   
However, as pointed out by Hoyer and  Peign{\'e}\cite{Hoyer} very recently, 
the ratio of $\psi '$ to $J / \psi$ photoproduction cross sections 
calculated in the dipole approximation is much smaller 
than the measured values\cite{ExpHERA} at HERA, when one 
uses typical non-relativistic quark models to obtain the charmonium wave 
functions.  
The formula used in the dipole approach\cite{Brodsky,FKS,Hoyer} 
is obtained by assuming that 
both transverse momentum of exchanged gluons and the transverse size of 
the $c \bar c$ pair are small.  
However, the non-relativistic quark model indicates that 
the typical size of $\psi '\sim 0.8 \mbox{fm}$ is twice as large as 
the radius of $J / \psi$.  
Hence, the validity of the small-size dipole approximation is questionable 
for $\psi ' (2S)$ and other mesons with the large radii.  
We may  expect considerable contributions from the scattering of the 
color dipole with a large transverse fluctuation, if the $\gamma$ and 
$c \bar c$ wave functions allow such a large separation 
between the quarks.  
Above argument motivates us to calculate the diffractive $J / \psi $ and 
$\psi '$ productions without resorting to the dipole approximation.  

\section{Calculation of diffractive heavy quarkonium production}

Let us briefly introduce the framework to calculate the diffractive 
heavy quarkonium production, which was developed in ref.~\cite{FKS}.  
We consider the forward heavy vector meson production 
$\gamma^{(*)} (q) + p(p) \to V (q) + p(p)$ shown 
in Fig.1 with the kinematical condition 
$W^2 \equiv (p+q)^2 \gg Q^2(=-q^2)$.  
In this small-$x$ kinematics, the lifetime of a color-singlet 
quark-antiquark fluctuation is 
large compared to the typical interaction time scale\cite{Text}.  
Then the production 
amplitude can be expressed as a convolution of 
the $q \bar q$-proton cross section with the wave functions of initial 
($\gamma$) and final ($V$) states.  
In terms of light-cone wave functions of the photon $\psi_\gamma$ and vector 
meson $\psi_V$, the production 
amplitude at $t=0$ is proportional to  
\begin{eqnarray}
{\cal A} (z) = 4 \pi \alpha_s\int d^2 b  \int d 
l_t^2  \frac{ f(x,l_t)}{l_t^4} \;  
\psi_\gamma (Q^2;z,b)^* \, [ 1 - J_0 (l_t \, b)] \, \psi_{V} (z,b) \; , 
\label{amp1}
\end{eqnarray}
where $z$ and $b$ are longitudinal momentum fraction of a quark and 
transverse distance between the $c \bar c$ pair, respectively.   
$x$ is the longitudinal momentum fraction of the gluon $x \approx (Q^2 + 
4 m^2) / W^2$ with $m$ being the heavy quark mass,  
and $l_t$ the gluon's transverse momentum.  
$f(x,l_t)$ denotes the unintegrated gluon number density of the proton, 
which contains a complicated non-perturbative gluon ladder.  
$\alpha_s$ and $J_0(bl_t)$ represent the QCD coupling constant and the 
ordinary Bessel function of the first kind.  
After multiplying suitable functions of $z$ and kinematical factors, and 
carrying out $z$-integration, we obtain both longitudinal and transverse 
differential cross sections for the heavy quarkonium production\cite{FKS};
\begin{eqnarray}
\left( \frac{ d \sigma } { dt} \right)_{t=0} 
\propto \left[ \frac{Q^2}{M_V^2} \left(\int_0^1 dz \, {z(1-z)}
 {\cal A}(z) \right)^2 
+ \left( \frac{m^2}{4 M_V^2}\int_0^1 \frac{dz} {z(1-z)} 
{\cal A}(z) \right)^2 \right] \; \; , 
\label{cross}
\end{eqnarray}
where $M_V$ is the mass of the heavy quarkonium.  

In eq.~(\ref{amp1}) the photon wave function $\psi_\gamma$ with 4-momentum 
square $Q^2$ is calculated perturbatively, while $\psi_V$ is a 
non-perturbative object.  
We shall use the non-relativistic potential model which successfully 
reproduces the properties of the heavy quarkonium to evaluate
$\psi_{V}$.  
In the following calculations, we use two types of the potential; 
the QCD motivated potential  
which essentially consists of linear and Coulomb potentials\cite{BT}, 
and a phenomenological logarithmic potential \cite{log}.  
Our procedure to evaluate the 
light-cone wave functions is in line with those of refs.~\cite{FKS,Hoyer}.
Solving the Schr{\"o}dinger equation with a potential $V(r)$, we obtain 
non-relativistic wave functions $\psi_V (\vec r)$, and also subsequently 
those in the 
momentum space $\psi_V (\vec k)$ through the Fourier transformation.  
We make a simple kinematical identification between  light-cone 
variables $(z, k_t)$ and $\vec k = (k_z, \vec k_t)$ as
\begin{eqnarray}
z = \frac{1}{2} \left( 1+ \frac{k_z}{\sqrt{\vec k^2 + m^2}} \right) \; ,
\label{momentum}
\end{eqnarray}
where $\pm k_t$ are the transverse momenta of the quarks.  
This relation gives the light-cone wave function;
\begin{eqnarray} 
\psi_V (z,k_t) =  \left({\frac{k_t^2 + m^2 }{4 [z(1-z)]^3}}\right)^{1/4} 
 \cdot 
\psi_V \left( |\vec k| = \sqrt{\frac{k_t^2+(2z -1)^2 m^2}{4 z (1-z)}}
\right) \; \; .
\label{wf}
\end{eqnarray}
Finally, performing the two-dimensional Fourier transformation in the 
$k_t$ space, we obtain the light-cone wave function 
$\psi_V (z,b)$.

If we take the infinitely heavy quark mass limit, where 
the Fermi motion of heavy quarks in the quarkonium can be entirely 
neglected, {{\it i.e}.} $\psi_{V}=\delta(z-1/2) \delta^{(2)} (b)$,  
the production cross section is simply related to the 
leptonic decay width of the heavy quarkonium\cite{Ryskin}.  
However, for the charmonium, there exist substantial Fermi motion 
corrections at small $Q^2$\cite{FKS,Ryskin2,Nemchik}.

Assuming that $(b \, l_t) \ll 1$ and keeping terms up to second order of
$(b \, l_t)$ \cite{FKS,Nemchik,Hoyer}, one can rewrite eq.~(\ref{amp1}) as
\begin{eqnarray}
{\cal A} (z) =  \alpha_s \,  x G(x, Q^2_{eff})\;  \left[ \pi \int d^2 b \, 
\psi_\gamma (Q^2;z,b)^* \, b^2 \,  \psi_{V} (z,b) \right] \; \; , 
\label{dipole}
\end{eqnarray}
where $Q^2_{eff} = z(1-z) Q^2 +  m^2$.    
In eq.~(\ref{dipole}) the unintegrated gluon density has been identified 
with the gluon distribution function measured in DIS via\cite{Text}
\begin{eqnarray*}
f(x,l_t)
\equiv {l_t^2}  \frac{ \partial \, x G(x,l_t^2)}{\partial \, l_t^2} \; .
\end{eqnarray*}
The approximation (\ref{dipole}) is often called the color 
dipole approximation.  
It gives a simple physical interpretation of this process, 
in which the scattering amplitude 
of the $q \bar q$ dipole with the target has a geometrical 
expression, $\sigma_{q \bar q} \propto b^2$.  
In addition, the factorized formula eq.~(\ref{dipole})
makes it possible to constrain the gluon distribution function of the 
proton from the diffractive vector meson production.  
Typical results for $J /\psi$ production cross section 
obtained by using eq.~(\ref{dipole}) are available in refs.~\cite{FKS,h}, 
and in agreement with the data.  
However, if we focus on $\psi '$, which is a radially excited 
$2S$-state of the charmonium, the calculated ratio of $\psi '(2S)$
to $J / \psi (1S)$ cross sections at the photoproduction point 
$Q^2 = 0$ is much smaller than the experimental data as pointed out 
in \cite{Hoyer}. The ratio calculated using eq.~(\ref{dipole}) is 
shown by  the dashed curve in Fig.2.   Although  
the theoretical prediction agrees with the data at larger $Q^2$, it has 
a factor 3 discrepancy at $Q^2 = 0$.  
We have tried to evaluate it with several potential models, 
but none of them can describe the data as far as the parameters of the 
potential are fixed to reproduce the charmonium properties\cite{h}.

Let us now examine carefully the dipole approximation which leads us 
to eq.~(\ref{dipole}).  
For the leptoproduction with large $Q^2$  eq.~(\ref{dipole}) works well, 
since the photon wave function projects 
out the $c \bar c$ fluctuation with the small transverse size.   
This approximation is also reasonable for the photoproduction of the 
$1S$ heavy quarkonium due to the small transverse size 
of the $J / \psi$.  
However, for the photoproduction of the radially excited 
vector mesons, it is not clear whether the small-$(b \, l_t)$ 
expansion of eq.~(\ref{dipole}) is valid.   
Since the size of $\psi ' (2S)$ turns out to be about 
$0.8 \mbox{fm}$, twice as large as that of $J / \psi (1S)$, 
in the typical non-relativistic quark model, the color dipole 
expansion may not work in this case.  
Namely, there may exist non-negligible contributions to the integral 
eq.~(\ref{amp1}) 
>from an overlap of the large-size $c \bar c$ configuration and 
the $\psi '$ wave function.  

In this work we explicitly evaluate eq.~(\ref{amp1}) 
instead of using the simplified formula (\ref{dipole}).  
To do this, we need the explicit form of the unintegrated 
gluon density $f (x, l_t)$, which is not precisely known.    
Here, we follow the prescription 
of refs.~\cite{Levin,Ryskin2} 
to carry out the $l_t$-integration in eq.~(\ref{amp1}) by relating 
$f(x,l_t)$ to the derivative of the gluon distribution;
\begin{eqnarray}
\int_0^{Q^2_{eff}} d l_t^2 \frac{f(x,l_t)}{l_t^2} A(l_t) &=& 
\int_0^{Q^2_0} dl_t^2 \frac{f(x, l_t)}{l_t^2} A(l_t) + 
\int_{Q^2_0}^{Q^2_{eff}} dl_t^2 \frac{f(x, l_t)}{l_t^2} A(l_t) \nonumber \\
&\sim& x G(x,Q^2_0) A(l_t=0) + 
\int_{Q^2_0}^{Q^2_{eff}} dl_t^2 \frac{\partial \, x G(x,l^2_t)} 
{\partial \, l_t^2} A(l_t)
\label{app}
\end{eqnarray}
where $A(l_t)$ is a given function of $l_t$ in eq.~(\ref{amp1}), 
and $Q_0^2$ plays a role of the infrared separation scale.  

\section{Results and discussions}

Using the CTEQ4LQ\cite{CTEQ} parametrization for the gluon distribution and 
taking $Q^2_0 = 0.8 \mbox{GeV}^2$, we obtain the result shown in Fig.2.  
Calculation without (with) the dipole approximation is depicted 
by the solid (dashed) curve.   For large $Q^2$ 
the difference between them becomes small as expected.  
However, around $Q^2 = 0$, the calculation of 
eq.~(\ref{amp1}) with eq.(\ref{app})  clearly larger than 
the result obtained by eq.~(\ref{dipole}) by factor about 2\cite{foot}, 
and shows a reasonable agreement with the experimental data\cite{ExpHERA}.   
We demonstrate the major 
improvement of the ratio is caused by the contribution from the large-size 
$q \bar q$ fluctuation.  
There may be several other mechanisms to enhance the $\psi '$ to $J /\psi$ 
ratio further, which will be discussed elsewhere\cite{h}.  

If we vary the value of $Q^2_0$ from $0.4 \mbox{GeV}^2$ to $1\mbox{GeV}^2$, 
resulting 
ratio changes within a few $\%$.  Our results are also found to be 
insensitive to the choice of the potential model.  
We show in Fig.3 results with GRV98LO\cite{GRV} parametrization for the gluon 
distribution function (thin-solid curve) in comparison with those with the 
CTEQ4LQ distribution (thick-solid curve).   
Calculated ratio is slightly modified due to the difference of the 
gluon distributions between CTEQ4 and GRV98 at low $Q^2$.  

For completeness, sensitivity of the results to $Q^2_{eff}$ is demonstrated 
in Fig.3, because several choices of 
$Q^2_{eff}$ are adopted in the literature\cite{FKS,Ryskin2,Nemchik}.  
In Fig.3 results with $Q^2_{eff} \to \infty$ as an extreme case 
are shown by the thick-dashed (CTEQ) and thin-dashed (GRV) curves, while those 
with $Q^2_{eff} = Q^2 / 4 + m^2$ by the solid curves.  
Although increasing the upper limit of the $l_t$-integration 
improves the result,   
significant enhancement at $Q^2 = 0$ comes from higher order 
terms of $b^2$ in eq.~(\ref{amp1}).

We also consider the production of $\Upsilon(1S,2S,3S)$-states.  One 
can naively expect that the transverse separation $b$ is small enough to 
ensure the validity of the dipole expansion for the bottom quark case.  
Results for $\Upsilon ' (2S) / \Upsilon (1S)$ 
and  $\Upsilon '(3S) / \Upsilon (1S)$ are shown in Fig.4.  They are consistent 
with those obtained in ref.~\cite{bottom} by the dipole formula.  
There are only slight differences between the calculations without (solid) 
and with (dashed) the dipole approximation  
even for the $\Upsilon '(3S)$ case, whose radius is about $0.6 \mbox{fm}$.

In conclusion, we have studied the diffractive photo- and 
leptoproduction of radially excited heavy quarkonia, and 
discussed the validity of the small-size color 
dipole approximation, which gives the smaller value of 
the $\psi '$ to $J / \psi$ ratio compared with the experimental data.    
It is pointed out that the contribution from the higher order terms of $b^2$ 
in the $q \bar q$-nucleon scattering amplitude, 
which is inherent in the original formula eq.~(\ref{amp1}),  
is important for the radially excited $\psi '$ case.  
We have demonstrated that the 
calculation of eq.~(\ref{amp1}) without the dipole approximation 
provides a much larger value for the elastic photoproduction ratio 
$d \sigma (\psi ' (2S)) / d \sigma (J / \psi (1S))$ 
than the dipole result, and is consistent with the HERA data.  
As claimed in ref.~\cite{Hoyer}, relativistic treatment of the 
charmonium system may be 
of some importance to construct the $\psi '$ wave function. 
However, what we have shown here is that, 
even within our non-relativistic framework where there are some 
ambiguities of choosing the kinematics {{\it e.g}.} in eq.~(\ref{momentum}), 
substantial improvement  for the $\psi '$ production can be achieved by 
evaluating eq.~(\ref{amp1}) instead of eq.~(\ref{dipole}).    
Our results suggest that a careful treatment is necessary  
to study the radially excited light vector meson leptoproduction, 
{{\it e.g}.}~$\rho ' (2S), \phi', \cdots$ at moderate $Q^2$. 
We have also discussed the $\Upsilon ' (2S,3S)$ production processes, and 
found that 
eq.~(\ref{dipole}) is a good approximation for the
$b\bar b$ system because of its heavier quark mass.

\acknowledgements

We thank Dr.~D.~Jido for useful discussions on the early stage of the 
present work.  K.S. and A.H. are supported by JSPS Research Fellowships 
for Young Scientists.  
J.A. and T. H. are partially supported by
JSPS grant No. 98360.

%

\newpage

\begin{figure}[htb]
\epsfxsize = 9cm   
\centerline{\epsfbox{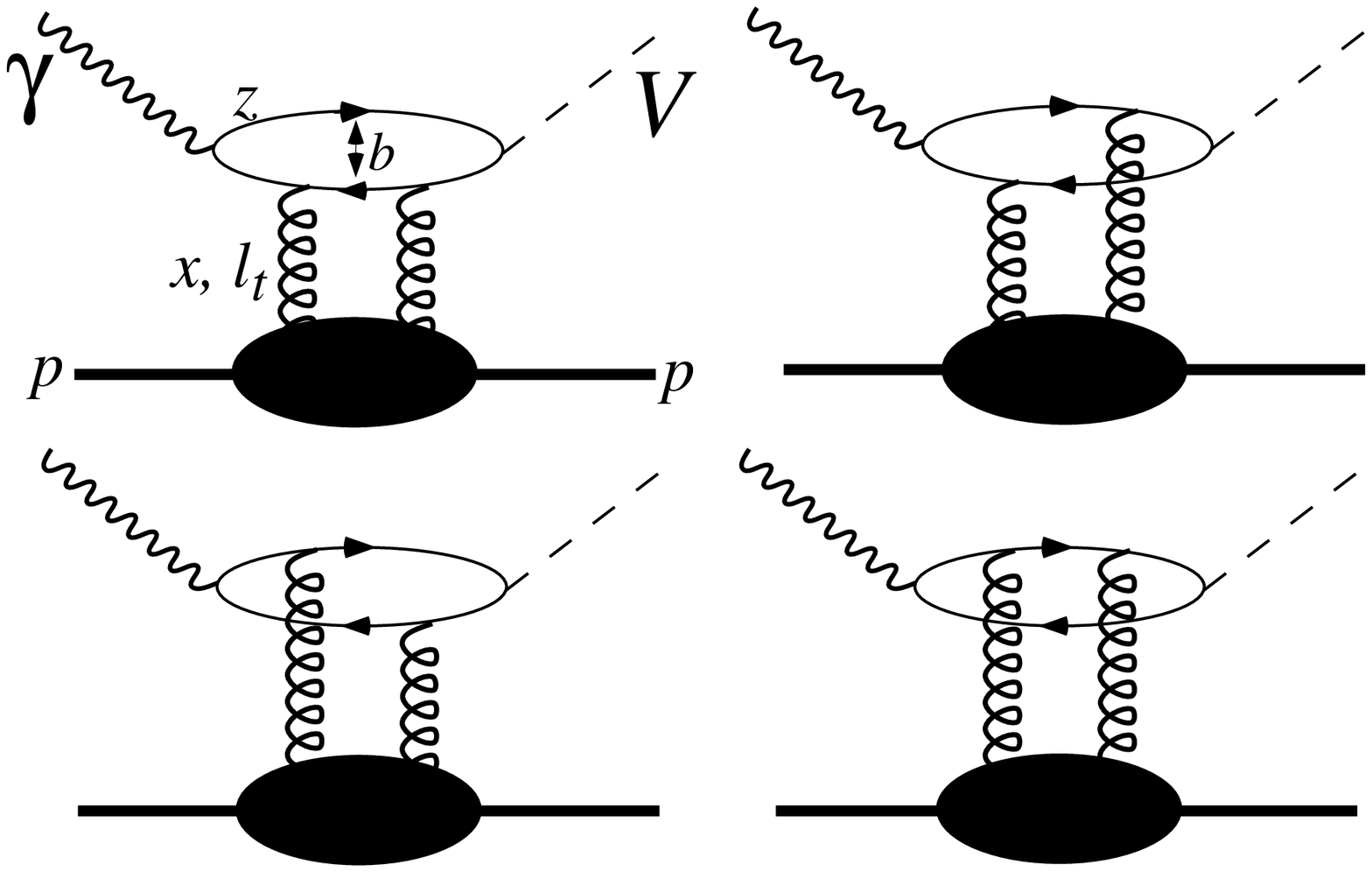}}
\caption{Diagrams contributing to the diffractive vector meson 
production. }

\vspace{1cm}

\epsfxsize = 9cm   
            \centerline{\epsfbox{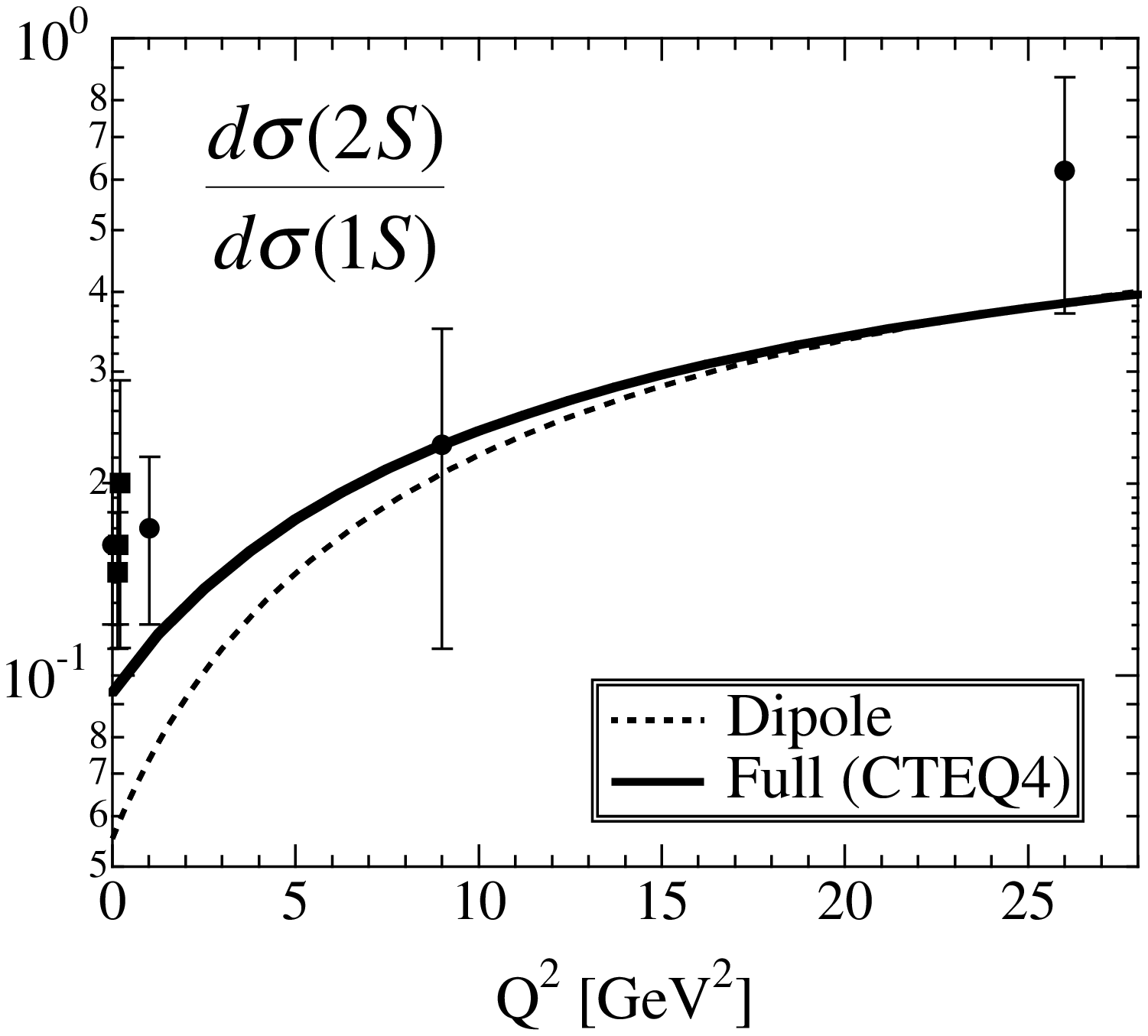}}
\caption{Ratio of $\psi'$ to  $J / \psi$ production cross sections at $W=
90 \mbox{GeV}$.  
Results obtained by the calculations with $Q^2_0 = 0.8 \mbox{GeV}^2$ 
and the color dipole approximation 
are depicted by the solid and dashed curves respectively. }

\vspace{1cm}

\epsfxsize = 9cm   
\centerline{\epsfbox{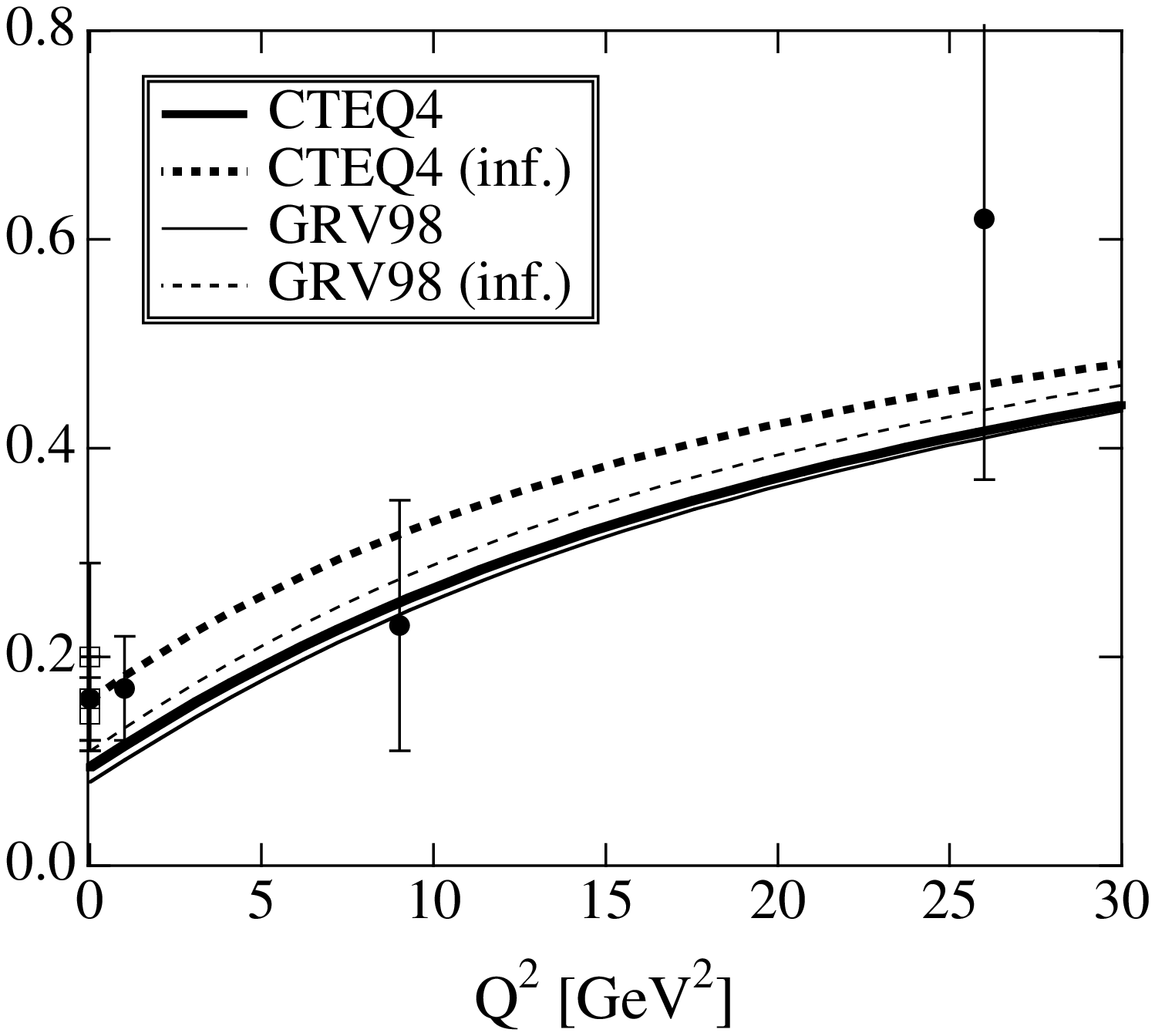}}
\caption{$G(x,Q^2)$ and $Q^2_{eff}$-dependence of the results with 
$Q^2_0 = 0.8 \mbox{GeV}^2$.  See text for details.}

\vspace{1cm}

\epsfxsize = 9 cm   
\centerline{\epsfbox{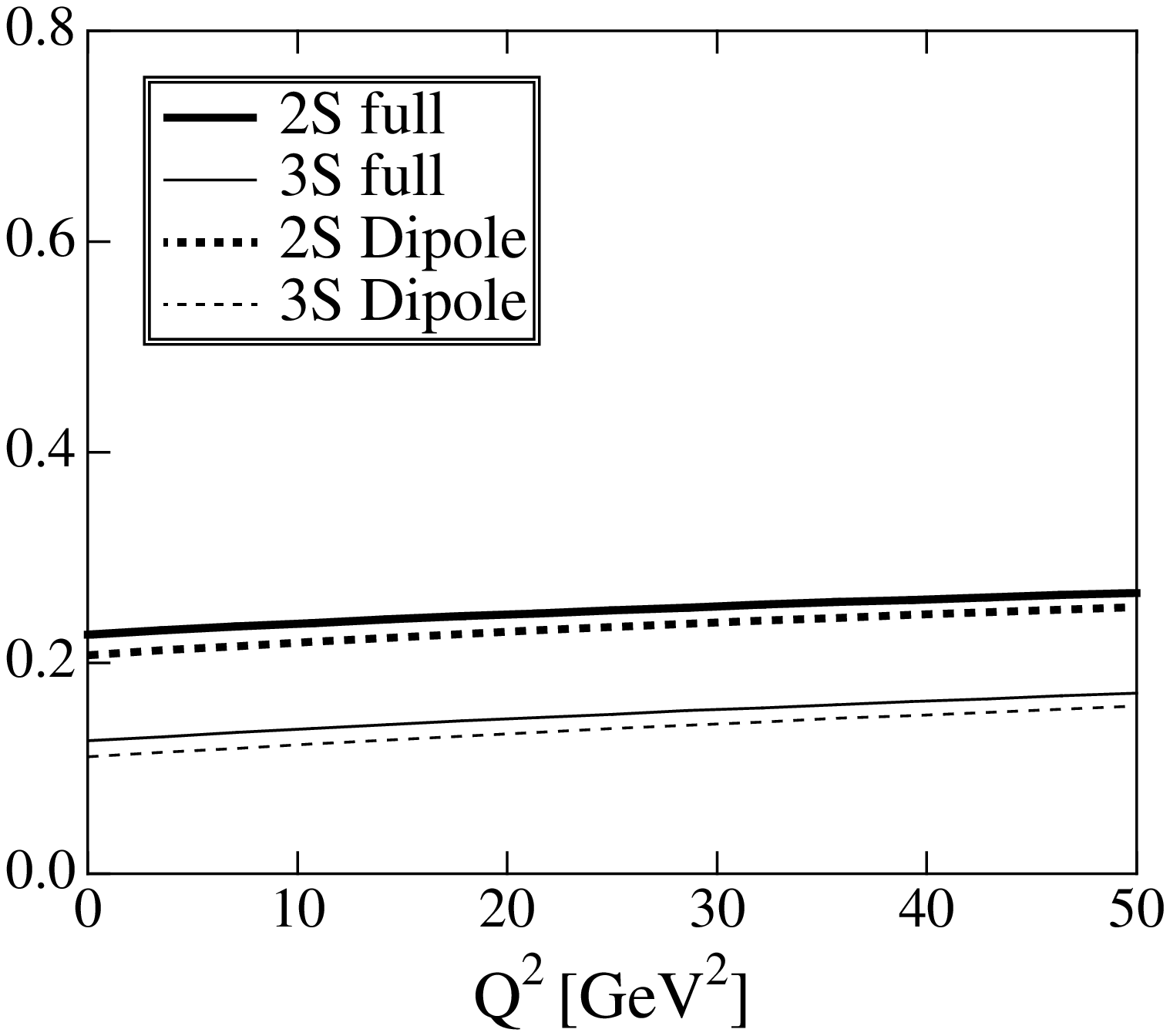}}
\caption{Ratios of $\Upsilon '(2S,3S)$ to $\Upsilon $ cross sections.  
Thick solid and dashed curves denote $d \sigma (\Upsilon '(2S))/d \sigma
(\Upsilon (1S))$ obtained by the full calculation and dipole approximation
respectively. Thin solid and dashed curves indicate the corresponding results 
for $\Upsilon '(3S) / \Upsilon(1S) $. }
\end{figure}


\begin{references}
\bibitem{Ryskin}M.~Ryskin, Z. Phys. {\bf C57}, 89 (1993) 
\bibitem{Brodsky}S. Brodsky {{\it et al}.}, Phys. Rev. {\bf D50}, 3134 (1994) 
\bibitem{FKS}
L. Frankfurt, W. Koepf and M. Strikman, Phys. Rev. {\bf D57}, 512 (1998)
\bibitem{Ryskin2}
M. Ryskin, R. Roberts, A. Martin, E. Levin, Z. Phys. {\bf C76}, 231 (1997)
\bibitem{Martin}
A. Martin, M. Ryskin, T. Teubner, Phys. Rev. {\bf D55}, 4329 (1997)
\bibitem{Nemchik}J. Nemchik {{\it et al}.}, JETP {\bf 86}, 1054 (1998) 
\bibitem{DL}A.~Donnachie and P.~V.~Landshoff, 
Phys.~ Lett.~{\bf B470}, 243 (1999)
\bibitem{Review}M. W{\"u}sthoff and A. Martin, J. Phys. {\bf G25}, 
R309 (1999) and references therein
\bibitem{Hoyer}P. Hoyer and S. Peign{\'e}, Phys. Rev. {\bf D61}, 031501(R) 
(2000)
\bibitem{ExpHERA}C.~Adolf {{\it et al}.} (H1 collaboration), 
Eur.~Phys.~J.~{\bf C10}, 373 (1999)
\bibitem{Text}J.~R.~Forshaw and D.~A.~Ross, 
{\it ``Quantum Chromodynamics and the Pomeron''},  
(Cambridge University Press, London, 1997)
\bibitem{BT}W.~Buchm{\"u}ller and S.-H.~Tye, Phys.~Rev.~{\bf D24}, 132 (1981)
\bibitem{log}C.~Quigg and J.L.~Rosner, Phys.~Lett.~{\bf B71}, 153 (1977)
\bibitem{h}A. Hayashigaki {{\it et al}.}, in preparation
\bibitem{Levin}E.~Levin {{\it et al}.}, Z. Phys. {\bf C74},  671 (1997)
\bibitem{CTEQ}H.L.~Lai {{\it et al}.}, Phys.~Rev.~{\bf D55}, 1280 (1997)
\bibitem{foot}We have also calculated eq.~(\ref{amp1}) 
using the `linear' approximation\cite{Martin,Levin} 
for the unintegrated gluon density, 
instead of the approximation adopted in eq.~(\ref{app}).  
Resulting $\psi '$ to $J /\psi$ ratio at $Q^2=0$ 
increases about $20 \%$ compared with the 
result shown in Fig.2, and comes closer to the data.  More realistic 
approximation for $f(x,l_t)$ 
may improve our result.  The result discussed in this paper 
should be considered as a lower bound of the ratio obtained by the 
calculation of eq.~(\ref{amp1}) without the dipole approximation.  
\bibitem{GRV}M.~Gl{\"u}ck, E.~Reya and A.~Vogt,
Eur.~Phys.~J.~{\bf C5}, 461 (1998)
\bibitem{bottom}L. Frankfurt, M. McDermott and M. Strikman, 
JHEP {\bf 9902}, 002 (1999)
\end{references}
\end{document}